\documentclass[aps,amsfonts,amssymb,preprint,eqsecnum,nofootinbib,superscriptaddress]{revtex4}
\usepackage{epsfig}
\usepackage{amsmath,amsfonts,bm}
\begin{document}

\count255=\time\divide\count255 by 60 \xdef\hourmin{\number\count255}
  \multiply\count255 by-60\advance\count255 by\time
 \xdef\hourmin{\hourmin:\ifnum\count255<10 0\fi\the\count255}

\newcommand{\Dslash}{D\hspace{-0.7em}{ }\slash\hspace{0.2em}}
\newcommand\<{\langle}
\renewcommand\>{\rangle}
\renewcommand\d{\partial}
\newcommand\LambdaQCD{\Lambda_{\textrm{QCD}}}
\newcommand\tr{\mathop{\mathrm{Tr}}}
\newcommand\+{\dagger}
\newcommand\g{g_5}

\newcommand{\xbf}[1]{\mbox{\boldmath $ #1 $}}

\title{A Higher-Derivative Lee-Wick Standard Model}

\author{Christopher D. Carone}
\email{cdcaro@wm.edu}

\affiliation{Department of Physics, College of William \& Mary,
Williamsburg, VA 23187-8795}

\author{Richard F. Lebed}
\email{Richard.Lebed@asu.edu}

\affiliation{Department of Physics, Arizona State University, Tempe,
AZ 85287-1504}

\date{November 2008}

\begin{abstract}
The Lee-Wick Standard Model assumes a minimal set of higher-derivative
quadratic terms that produce a negative-norm partner for each Standard
Model particle.  Here we introduce additional terms of one higher
order in the derivative expansion that give each Standard Model
particle two Lee-Wick partners: one with negative and one with
positive norm.  These states collectively cancel unwanted quadratic
divergences and resolve the hierarchy problem as in the minimal
theory.  We show how this next-to-minimal higher-derivative theory may
be reformulated via an auxiliary field approach and written as a
Lagrangian with interactions of dimension four or less.  This mapping
provides a convenient framework for studies of the formal and
phenomenological properties of the theory.
\end{abstract}


\maketitle

\section{Introduction} \label{intro}
Extensions of the Standard Model (SM) generally involve mass scales
that are much higher than the scale of electroweak symmetry breaking.
If one views the SM as a low-energy effective theory, then the Higgs
boson squared mass $m_h^2$ receives radiative corrections that grow
quadratically with the cutoff.  This leads to the hierarchy problem: A
large separation of scales requires an extremely close cancellation
between the bare Higgs boson mass and the cutoff-dependent loop
corrections.  Within the low-energy effective theory, such a fine
tuning has no natural explanation.

Solutions to the hierarchy problem can be grouped into three broad
categories, distinguished by their assumptions: (1) models that assume
fine tuning to be extreme and present, but natural from the point of
view of the string landscape, as in split-supersymmetric
models~\cite{splitsusy}; (2) models that assume fine tuning is not
extreme since no high mass scales are present, as in scenarios with
large extra dimensions and a low Planck scale~\cite{ADD}; (3) models
that assume fine tuning is not extreme, even when high mass scales are
present, because new physics just above the electroweak scale modifies
the ultraviolet divergence of $m_h^2$ from quadratic to logarithmic.
The Minimal Supersymmetric Standard Model (MSSM) is perhaps the most
famous example of a model in the last category: Each SM particle has a
supersymmetric partner with the same gauge quantum numbers but
opposite spin statistics.  As fermion and boson loops enter with
opposite relative signs, quadratic divergences cancel between Feynman
loop diagrams when both particles and their associated superpartners
are taken into account.

A similar cancellation is achieved in the Lee-Wick Standard Model
(LWSM)~\cite{GOW}, which has recently been proposed as a theory that
solves the hierarchy problem.  Each SM particle possesses a Lee-Wick
(LW) partner~\cite{oldLW} with the {\em same} spin statistics, but
with opposite-sign quadratic terms.  Since the propagators of ordinary
and LW particles differ in overall sign, quadratic divergences cancel
between pairs of diagrams.  A LW partner for a given field arises via
the inclusion of a higher-derivative (HD) kinetic term which generates
an additional pole in the associated two-point function.  As reviewed
below, the HD Lagrangian can be recast, using auxiliary fields, as a
dimension-four Lagrangian that includes partner fields with
``wrong-sign" quadratic terms~\cite{GOW}.  The cancellation of
divergences in this formulation of the theory occurs because HD terms
in the original Lagrangian cause propagators to fall off more quickly
with momentum, so that loop diagrams become less divergent.

While LW particles have wrong-sign kinetic and mass terms (like
Pauli-Villars regulators) it is nonetheless believed consistent to
treat them as physical particles.  Neither the LWSM~\cite{GOW}, in
which all the LW states can decay, nor the $O(N)$ LW model at large
$N$~\cite{causal} violates causality at a macroscopic level.
Moreover, studies of longitudinal gauge-boson scattering in the LWSM
indicate that unitarity is not violated provided the HD theory can be
mapped to a Lagrangian with interactions of dimension four or
less~\cite{unitary}.  Taking these observations into account, a number
of authors have begun to explore the
phenomenology~\cite{carleb,LWpheno} and cosmology~\cite{LWcosmo} of LW
extensions of the SM\@.  These studies have assumed the minimal
theory, in which the lowest-order HD term for each field is included,
and precisely one LW partner accompanies each SM particle.

While the minimal scenario is the simplest to study, one may wonder
whether the inclusion of a single HD term, and {\em exactly\/} no
others of higher order, represents a natural state of affairs.  In
this paper we explore a next-to-minimal scenario that includes HD
terms of the next order in a derivative expansion, leading to {\em
two\/} partners for each SM particle.  Our immediate focus is a
technical one: What is the generalization of the auxiliary field (AF)
formulation introduced in the minimal theory~\cite{GOW}, and what form
of the HD Lagrangian leads to an auxiliary field theory with
interactions of dimension four or less?  We address this question in a
non-Abelian gauge theory with fermions and complex scalars, so that
our results can be immediately applied to the SM\@.  Interestingly,
one of the two new LW partners for each SM particle is ordinary (with
correct-sign quadratic terms), suggesting that collider signatures and
experimental limits on this theory can be qualitatively different from
the minimal version.  Our results suggest that there is no impediment,
in principle, to constructing similar theories with additional LW
states via the inclusion of appropriate interactions that are of yet
higher order in the number of derivatives.

We note that previous work~\cite{JKL,JKL2} extensively studies a 
particular $O(p^6)$ form for a HD scalar Lagrangian, in which $O(p^4)$ terms 
are absent and gauge couplings are omitted.   In particular, this work develops 
a strongly-interacting Higgs sector that tames ultraviolet corrections and can 
be studied on the lattice.  Reference~\cite{JKL} represents pioneering early work 
on the consistency of $O(p^6)$ scalar theories.  By contrast, the thrust here is to 
study the duality between more general HD theories with $O(p^6)$ terms and equivalent 
theories with operators of dimension four or less, not only in the Higgs sector 
but including all SM particles, with an eye toward future phenomenological studies.

This paper is organized as follows.  In the next section we review the
LW idea in a simple scalar field theory and show how the AF
formulation is applied when HD terms of next-to-lowest order are
present.  In Section~\ref{sec:ym} we extend our approach to
non-Abelian gauge theories, focusing on the pure gauge sector; in
Section~\ref{sec:fermions} we show how fermions are included in the
theory.  In Section~\ref{sec:higgs} we discuss the Higgs sector of the
theory.  In Section~\ref{divkill} we discuss the cancellation of one-loop 
quadratic divergences in an SU($N_c$) gauge theory with complex scalars and chiral 
fermions. In Section~\ref{sec:concl} we summarize our conclusions.

\section{A Scalar Example}\label{sec:scalarex}

Let us begin by reviewing the formulation of a LW theory of a real
scalar field.  The simplest HD Lagrangian is given by
\begin{equation}
{\cal L}_{{\rm HD}}=-\frac{1}{2} \hat{\phi} \,\Box\, \hat{\phi}
- \frac{1}{2M^2} \hat{\phi} \,\Box^2 \hat{\phi}
-\frac{1}{2} m_\phi^2 \hat{\phi}^2 +{\cal L}_{{\rm int}}(\hat{\phi})
\, ,
\label{eq:toyhd}
\end{equation}
where the last term represents interactions. The HD term leads to an
additional pole in the $\hat{\phi}$ two-point function near the mass
$M$, which corresponds to the LW partner of the usual state with mass
eigenvalue near $m_\phi$.  The HD term also assures high-momentum
falloff of the $\hat{\phi}$ propagator as $1/p^4$, improving the
convergence of $\hat{\phi}$ loop diagrams.  Following the approach of
Ref.~\cite{GOW}, one observes that Eq.~(\ref{eq:toyhd}) is equivalent
to a Lagrangian including an auxiliary field, $\tilde{\phi}$ and no
higher-derivative interactions:
\begin{equation}
{\cal L}_{{\rm AF}}= -\frac{1}{2} \hat{\phi} \,\Box\, \hat{\phi}
-\frac{1}{2} m_\phi^2 \hat{\phi}^2 - \tilde{\phi} \,\Box\, \hat{\phi}
+ \frac{1}{2} M^2 \tilde{\phi}^2 +{\cal L}_{{\rm int}}(\hat{\phi})
\, .
\label{eq:toyaf}
\end{equation}
The $\tilde{\phi}$ equation of motion (EOM) is
\begin{equation}
\tilde{\phi} = \frac{1}{M^2} \Box\,\hat{\phi} \, ,
\end{equation}
which, upon substitution into Eq.~(\ref{eq:toyaf}), reproduces the
original Lagrangian of Eq.~(\ref{eq:toyhd}).  The kinetic terms in
Eq.~(\ref{eq:toyaf}) can be diagonalized via the substitution
\begin{equation}
\hat{\phi}=\phi-\tilde{\phi} \, ,
\label{eq:shift}
\end{equation}
yielding
\begin{equation}
{\cal L}=-\frac{1}{2} \phi \,\Box\, \phi+\frac{1}{2} \tilde{\phi}
\,\Box\, \tilde{\phi} -\frac{1}{2} m_\phi^2
(\phi-\tilde{\phi})^2+\frac{1}{2} M^2 \tilde{\phi}^2 +{\cal L}_{{\rm
int}}(\phi-\tilde{\phi}) \, .
\end{equation}
The scalar mass matrix can be diagonalized without affecting the form
of the kinetic terms via a symplectic transformation:
\begin{equation}
\left(\begin{array}{c} \phi \\ \tilde{\phi}\end{array}\right) =
\left(\begin{array}{cc} \cosh\theta & \sinh\theta \\ \sinh\theta &
\cosh\theta \end{array}\right) \left(\begin{array}{c} \phi_0 \\
\tilde{\phi}_0 \end{array}
\right) \, ,
\end{equation}
where the subscript $0$ indicates a mass eigenstate; one finds
\begin{equation}
\tanh 2\theta = \frac{-2 m_\phi^2}{M^2-2 m_\phi^2}  \, .
\end{equation}
The final Lagrangian takes the form
\begin{equation}
{\cal L}_{{\rm LW}} = -\frac{1}{2} \phi_0 \,\Box\, \phi_0+\frac{1}{2}
\tilde{\phi}_0 \,\Box\, \tilde{\phi}_0 -\frac{1}{2} m_0^2 \phi_0^2
+\frac{1}{2} M_0^2 \tilde{\phi}_0^2 +{\cal L}_{{\rm
int}}[e^{-\theta}(\phi_0-\tilde{\phi}_0)] \, ,
\label{eq:toylw}
\end{equation}
where $m_0$ and $M_0$ are the mass eigenvalues, and the factor of
$e^{-\theta}$ can be absorbed into redefinitions of the couplings.
The opposite-sign $\phi_0$ and $\tilde{\phi}_0$ propagators following
from the quadratic terms in Eq.~(\ref{eq:toylw}), together with the
specific relationship between the $\phi_0$ and $\tilde{\phi}_0$
couplings in ${\cal L}_{{\rm int}}$, assures the cancellation of
quadratic divergences, as is shown explicitly in Ref.~\cite{GOW}.

Indicating by $N$ the number of physical poles in the
$\hat{\phi}$ propagator, let us refer to the minimal example just
considered as an $N=2$ theory.  An $N \! = \! 3$ model corresponds to
a HD Lagrangian of the general form
\begin{equation}
{\cal L}_{{\rm HD}}^{N=3}=-\frac{1}{2} \hat{\phi} \,\Box\, \hat{\phi}
- \frac{1}{2M_1^2} \hat{\phi} \,\Box^2 \hat{\phi}- \frac{1}{2M_2^4}
\hat{\phi} \,\Box^3 \hat{\phi} -\frac{1}{2} m_\phi^2 \hat{\phi}^2
+{\cal L}_{{\rm int}}(\hat{\phi}) \, ,
\label{eq:toyhd3}
\end{equation}
where $M_1$ and $M_2$ are the LW mass scales, which we assume are
comparable.  The restriction that the $\hat{\phi}$ propagator has
three physical poles restricts the values of $m_\phi^2$, $M_1^2$ and
$M_2^2$, so that it is possible to map Eq.~(\ref{eq:toyhd3}) to a
Lagrangian of the form
\begin{equation}
{\cal L}_{{\rm LW}}^{N=3} = \sum_{i=1}^3 c_i \left[-\frac{1}{2}
\phi^{(i)}(\Box + m_i^2) \phi^{(i)}\right] +{\cal L}_{{\rm
int}}(\{\phi^{(i)}\}) \, ,
\label{eq:lw3}
\end{equation}
where the $c_i = 1 \mbox{ or } -1$, and the $m_i^2$ are positive.  The
missing link that connects Eq.~(\ref{eq:toyhd3}) to (\ref{eq:lw3}) is
an AF Lagrangian, analogous to Eq.~(\ref{eq:toyaf}) in the $N=2$
theory, and appropriate field redefinitions, analogous to
Eq.~(\ref{eq:shift}).  Let us first examine the special case where
$m_\phi=0$ [which corresponds to $m_1=0$ in Eq.~(\ref{eq:lw3})] before
stating the general result.  The desired AF Lagrangian involves two
new scalar fields, $\chi$ and $\psi$:
\begin{eqnarray}
{\cal L}_{\rm AF} = -\frac 1 2 \hat \phi \, \Box \, \hat \phi - \chi
\, \Box \, \hat \phi + m_2 m_3 \, \chi \psi - \frac 1 2 \psi \, \Box
\, \psi - \frac 1 2 (m_2^2 + m_3^2) \, \psi^2 + {\cal L}_{\rm int}
(\hat \phi) \, .
\label{eq:afm10}
\end{eqnarray}
Like the field $\tilde{\phi}$ in the $N=2$ theory, $\chi$ is an
auxiliary field; since it occurs linearly in Eq.~(\ref{eq:afm10}), its
EOM imposes a constraint that is exact at the quantum level:
\begin{equation}
\psi = \frac{1}{m_2 m_3} \, \Box \, \hat \phi \, .
\label{eq:cononpsi}
\end{equation}
Substituting Eq.~(\ref{eq:cononpsi}) into Eq.~(\ref{eq:afm10}), one
obtains
\begin{equation}
{\cal L}_{{\rm HD}}=-\frac{1}{2}\hat\phi\,\Box\,\hat\phi -\frac{1}{2}
\left(\frac{m_2^2+m_3^2}{m_2^2 m_3^2}\right)
\hat\phi \,\Box^2 \hat\phi -\frac{1}{2}
\left(\frac{1}{m_2^2 m_3^2}\right) \hat \phi \, \Box^3 \hat \phi
+ {\cal L}_{{\rm int}}(\hat\phi) \, ,
\label{eq:hdm10}
\end{equation}
which factorizes as
\begin{equation}
{\cal L}_{\rm HD} = -\frac{1}{2m_2^2 m_3^2} \hat \phi \, \Box \,
(\Box + m_2^2) (\Box + m_3^2) \, \hat \phi + {\cal L}_{{\rm int}}
(\hat\phi) \, , \label{LHDm10}
\end{equation}
and from which one identifies $m_\phi \! = \! 0$, $M_1^2 \! = \! m_2^2
m_3^2/(m_2^2+m_3^2)$ and $M_2^4 \! = \! m_2^2 m_3^2$ upon comparison
with Eq.~(\ref{eq:toyhd3}).

Showing next that the AF Lagrangian can also be written in the form of
Eq.~(\ref{eq:lw3}) is a simple matter of linear algebra.  Taking $m_2$
to be the lighter LW state and substituting the field redefinitions
\begin{eqnarray}
\hat \phi & = & \phi^{(1)} - \frac{m_3}{(m_3^2-m_2^2)^{1/2}}
\phi^{(2)}  + \frac{m_2}{(m_3^2-m_2^2)^{1/2}} \phi^{(3)} \, ,
\label{phidef} \\
\chi & = & \frac{1}{(m_3^2-m_2^2)^{1/2}} \left[ m_3 \phi^{(2)} - m_2
\phi^{(3)} \right] \, , \label{chidef} \\
\psi & = & \frac{1}{(m_3^2-m_2^2)^{1/2}} \left[ m_2 \phi^{(2)} - m_3
\phi^{(3)} \right] \, , \label{psidef}
\end{eqnarray}
into Eq.~(\ref{eq:afm10}), one obtains
\begin{equation}
{\cal L} = -\frac{1}{2} \phi^{(1)} \, \Box \, \phi^{(1)} + \frac{1}{2}
\phi^{(2)} ( \Box\, + \, m_2^2 ) \, \phi^{(2)} -\frac{1}{2} \phi^{(3)}
(\Box\, + \, m_3^2) \, \phi^{(3)} + {\cal L}_{{\rm
int}}( \hat{\phi} ) \, .
\label{eq:m10res}
\end{equation}

As with Eq.~(\ref{eq:shift}) in the $N \! = \! 2$ theory,
Eq.~(\ref{phidef}) leads to a very specific form for
the interaction terms in Eq.~(\ref{eq:m10res}).  We find
that there is no finite field redefinition that takes the AF
Lagrangian Eq.~(\ref{eq:afm10}) to the LW form Eq.~(\ref{eq:m10res}) 
for $m_2=m_3$, so we do not consider that possibility further.

For completeness, we exhibit the results for $m_\phi$ (and $m_1$)
non-zero.  The AF Lagrangian is given by
\begin{eqnarray}
 {\cal L}_{\rm AF} & = & \frac{1}{\eta_1}\left[-\frac 1 2 \hat \phi \,
( \Box + m_1^2 ) \hat \phi - \chi ( \Box + m_1^2 ) \hat \phi +
(m_3^2 - m_1^2)^{1/2} (m_2^2 - m_1^2)^{1/2} \chi \psi \right.
\nonumber \\ & &\left. - \frac 1 2 \psi \, \Box \psi - \frac 1 2
( m_2^2 + m_3^2 - m_1^2) \psi^2 \right] +
{\cal L}_{\rm int}(\hat \phi) \, ,
\label{eq:afstotal}
\end{eqnarray}
where $\eta_1 \! \equiv \! (m_1^2 m_2^2 + m_1^2 m_3^2 + m_2^2 m_3^2) /
(m_2^2 - m_1^2) (m_3^2 - m_1^2)$.  Varying Eq.~(\ref{eq:afstotal})
with respect to auxiliary field $\chi$ generalizes the EOM 
Eq.~(\ref{eq:cononpsi}) to
\begin{equation}
\psi = \frac{1}{(m_2^2 - m_1^2)^{1/2} (m_3^2 - m_1^2)^{1/2}} \,
(\Box + m_1^2) \, \hat \phi \, ,
\end{equation}
which, when substituted back into Eq.~(\ref{eq:afstotal}), yields
\begin{equation}
{\cal L}_{\rm HD} = -\frac{1}{2\Lambda^4} \hat \phi \, (\Box +
m_1^2) (\Box + m_2^2) (\Box + m_3^2) \, \hat \phi \, ,
\label{eq:hdfactored}
\end{equation}
where
\begin{equation}
\Lambda^4 \equiv m_1^2 m_2^2 + m_1^2 m_3^2 + m_2^2 m_3^2 \, .
\label{eq:lamsq}
\end{equation}
Equation~(\ref{eq:hdfactored}) is equivalent to the HD Lagrangian in
Eq.~(\ref{eq:toyhd3}) with the identifications
\begin{eqnarray}
m_\phi^2 & = & (m_1^2 m_2^2 m_3^2)/\Lambda^4 \, , \label{eq:msqds1}\\
M_1^2 & = & \Lambda^4/(m_1^2+m_2^2+m_3^2) \, , \label{eq:msqds2}\\
M_2^2 & = & \Lambda^2 \, .\label{eq:msqds3}
\end{eqnarray}
On the other hand, one can obtain the canonical LW form,
Eq.~(\ref{eq:lw3}) with $c_1 \! = \! -c_2 \! = \! c_3 \! = \! 1$,
from Eq.~(\ref{eq:afstotal}) by the field redefinitions
\begin{eqnarray}
\hat \phi & = & \sqrt{\eta_1} \, \phi^{(1)} \! - \sqrt{-\eta_2} \,
\phi^{(2)} + \sqrt{\eta_3} \, \phi^{(3)}  \, , \label{eq:redef1}\\
\chi & = & \sqrt{-\eta_2} \, \phi^{(2)} \! - \sqrt{\eta_3} \,
\phi^{(3)} \, , \label{eq:redef2}\\
\psi & = & \sqrt{\eta_3} \, \phi^{(2)} \! - \sqrt{-\eta_2} \,
\phi^{(3)} \, , \label{eq:redef3}
\end{eqnarray}
where the parameters $\eta_i$ are defined by
\begin{eqnarray}
\eta_1 &\equiv & \frac{\Lambda^4}{(m_2^2-m_1^2)(m_3^2-m_1^2)} \, ,
\label{eq:etadef1} \\
\eta_2 &\equiv & \frac{\Lambda^4}{(m_1^2-m_2^2)(m_3^2-m_2^2)} \, ,
\label{eq:etadef2}\\
\eta_3 &\equiv & \frac{\Lambda^4}{(m_1^2-m_3^2)(m_2^2-m_3^2)} \, .
\label{eq:etadef3}
\end{eqnarray}
Noting, for example, that $\eta_1 \! = \! 1$ when $m_1 \! = \! 0$, one
sees that Eqs.~(\ref{phidef})--(\ref{psidef}) immediately follow in
this case.  As before, we assume $m_3 > m_2 > m_1$, so that ${\rm
sign}(\eta_i) \!  = \!  (-1)^{i+1}$.  The remarkable algebraic
simplifications that occur in converting the AF Lagrangian are a
consequence of simple sum rules that are satisfied by the $\eta_i$:
\begin{equation} 
\sum_{i=1}^3 m_i^{2n} \,\eta_i =0  \ \ (n=0,1),
\label{eq:sr1}
\end{equation}
\begin{equation}
\sum_{i=1}^3 m_i^{2n} \,\eta_i =\Lambda^4  \ \ (n=2),
\label{eq:sr2}
\end{equation}
\begin{equation}
m_1^2 m_2^2 \eta_3 + m_2^2 m_3^2 \eta_1 + m_3^2 m_1^2 \eta_2 =
\Lambda^4 \, .
\label{eq:sr3}
\end{equation}
Our $\eta_i$ parameters are equivalent to those introduced by Pais and
Uhlenbeck~\cite{pu} (which we call $\eta^{\rm PU}_i$) to describe
purely quantum-mechanical theories with HD Lagrangians analogous to
those used here.  The mapping
\begin{equation} 
\eta_i = \frac{m_i^4 \Lambda^{2N-2}}{\Pi_j m^2_j} \, \eta^{\rm PU}_{i}
\end{equation}
converts the sum rules of Ref.~\cite{pu} into Eqs.~(\ref{eq:sr1}) and
(\ref{eq:sr3}) for the case $N \! = \! 3$, while Eq.~(\ref{eq:sr2}) is
linearly dependent on the others.

The interaction terms in the general $N=3$ theory are functions of
$\hat{\phi}$.  Following from Eq.~(\ref{eq:redef1}),
\begin{equation}
{\cal L}_{{\rm int}}(\hat{\phi}) \equiv {\cal L}_{{\rm int}}
\left(\sqrt{\eta_1} \, \phi^{(1)} \! -
\sqrt{-\eta_2} \, \phi^{(2)} + \sqrt{\eta_3} \, \phi^{(3)}\right) \, .
\label{eq:sints}
\end{equation}
The restriction on the form of the couplings imposed by
Eq.~(\ref{eq:sints}) is necessary for the cancellation of divergences.
This fact is illustrated in the following simple example: Let ${\cal
L}_{{\rm int}}(\hat{\phi})=\lambda \hat{\phi}^4/4!$, or equivalently,
\begin{equation}
{\cal L}_{{\rm int}}(\hat{\phi})=\frac{\lambda}{4!} \sum_{ijkl} 
\sqrt{|\eta_i \eta_j \eta_k \eta_l|}
\phi^{(i)}\phi^{(j)}\phi^{(k)}\phi^{(l)} \,\,\,.
\end{equation}
The self-energy for $\phi^{(1)}$ (corresponding to the state that is
present when the LW particles are decoupled) is given by
\begin{equation}
\Pi(p^2) = \lambda \eta_1 \int \frac{d^4 p}{(2 \pi)^4} \sum_k 
\left[\frac{(-1)^{k+1} \, i}{p^2-m_k^2}\right] |\eta_k| \,\,\,,
\end{equation}
where the factor $(-1)^{k+1}$ yields the appropriate overall sign for
each scalar propagator.  Using the fact that $(-1)^{k+1} |\eta_k| =
\eta_k$ and formally expanding the integrand, one finds
\begin{equation}
\Pi(p^2) = i \, \lambda \eta_1 \int \frac{d^4 p}{(2 \pi)^4} \sum_k 
\left(\frac{\eta_k}{p^2} + \frac{\eta_k m_k^2}{p^4} +
\frac{\eta_k m_k^4}{p^6} + \cdots \right) \,\,\,.
\end{equation}
The first two terms vanish as a consequence of the $n=0$ and $1$ sum
rules, Eq.~(\ref{eq:sr1}), respectively; these terms would otherwise
be quadratically and logarithmically divergent, respectively.
Although the interactions in the LW form of the $N=3$ theory are more
complicated than in the $N=2$ case, the sum rules satisfied by the
$\eta_i$ always provide the necessary algebraic miracles that cancel
the leading divergences in the theory\footnote{Despite this example,
$N>2$ LWSMs are not finite theories, but remain logarithmically
divergent, as can be shown by a generalization of the power-counting
argument given in Ref.~\cite{GOW}.}.

\section{Pure Yang-Mills Theory}\label{sec:ym}

We now generalize the approach of the previous section to a pure
Yang-Mills theory. The next-to-leading-order HD Lagrangian reads
\begin{equation}
{\cal L}_{\rm HD} = -\frac 1 2 \, {\rm Tr} \, \hat F_{\mu \nu} \hat
F^{\mu \nu} - \left( \frac{1}{m_2^2} \! + \! \frac{1}{m_3^2} \right)
{\rm Tr} \hat F_{\mu \nu} \hat D^\mu \hat D_\alpha \hat F^{\alpha \nu}
- \frac{1}{m_2^2 m_3^2} \, {\rm Tr} \hat F_{\mu \nu} \hat D^\mu \hat
D_\alpha \hat D^{[\alpha} \hat D_\beta \hat F^{\beta \nu ]} \, ,
\label{YMHD}
\end{equation}
where the superscript brackets indicate antisymmetrization of just the
first and last indices:
\begin{equation}
X^{[\alpha_1 \alpha_2 \cdots \alpha_{N-1} \alpha_N]} \equiv
X^{\alpha_1 \alpha_2 \cdots \alpha_{N-1} \alpha_N} -
X^{\alpha_N \alpha_2 \cdots \alpha_{N-1} \alpha_1} \, .
\end{equation}
Equation~(\ref{YMHD}) can be written in the elegant factorized form
\begin{equation}
{\cal L}_{\rm HD} = {\rm Tr} \, \hat F_{\mu \nu} \! \left( \frac 1 2
\, g^\mu{}_\alpha + \frac{\hat D^\mu \hat D_\alpha}{m_2^2} \right)
\left[ \left( \frac 1 2 \, g^\nu{}_\beta +
\frac{\hat D^\nu \hat D_\beta}{m_3^2} \right) g^\alpha{}_\lambda -
(\alpha \leftrightarrow \nu) \right] \hat F^{\beta \lambda} \, .
\end{equation}
The field strength $\hat F$, and the covariant derivative $\hat
D$ acting upon a field $X$ transforming in the adjoint
representation of the gauge group, are defined in the usual manner:
\begin{eqnarray}
\hat F^{\mu \nu} & \equiv & \partial^\mu \hat A^\nu - \partial^\nu
\hat A^\mu - ig \,[ \hat A^\mu , \hat A^\nu ] \, , \label{fs} \\
\hat D^\mu X & \equiv & \partial^\mu X - ig \,[ \hat A^\mu , X ] \, .
\label{cov}
\end{eqnarray}
This HD Lagrangian may be obtained from the equivalent Lagrangian
\begin{eqnarray}
{\cal L}_{\rm YM} & = & -\frac 1 2 \, {\rm Tr} \, \hat F_{\mu \nu}
\hat F^{\mu \nu} - {\rm Tr} \, \hat F^{\mu \nu} \! ( \hat D_\mu
\chi_\nu - \hat D_\nu \chi_\mu ) - \frac 1 2 \, {\rm Tr} \,
( \hat D_\mu \omega_\nu - \hat D_\nu \omega_\mu )^2 \nonumber \\
& & - 2 m_2 m_3 \, {\rm Tr} \, \chi_\mu \omega^\nu + (m_2^2 + m_3^2)
\, {\rm Tr} \, \omega_\mu \omega^\mu \, , \label{YMaux}
\end{eqnarray}
where the new fields $\chi$ and $\omega$ transform in the adjoint
representation.  Integration by parts on the second term leads to a
form for ${\cal L}_{\rm YM}$ in which no derivatives on $\chi$ appear,
making it an auxiliary field; since $\chi$ appears linearly in ${\cal
L}_{\rm YM}$, it is also a Lagrange multiplier.  The constraint
imposed by its EOM,
\begin{equation}
\hat D_\nu \hat F^{\nu \mu} -  m_2 m_3 \, \omega^\mu = 0 \, ,
\label{eq:givesomega}
\end{equation}
is exact at the quantum level.  Using Eq.~(\ref{eq:givesomega}) to
eliminate $\omega^\mu$ from Eq.~(\ref{YMaux}), one finds that the
terms proportional to $\chi$ cancel, and that the remaining terms
reduce to the HD Lagrangian, Eq.~(\ref{YMHD}).

In order to obtain a Lagrangian in the LW form, we rewrite the three
fields $\hat A$, $\chi$ and $\omega$ in terms of three new fields
$A_{1,2,3}$:
\begin{eqnarray}
A_1^\mu & \equiv & \hat A^\mu + \chi^\mu \, , \nonumber \\
A_2^\mu & \equiv & \sqrt{-\frac{\eta_2}{\eta_1}} \chi^\mu -
\sqrt{\frac{\eta_3}{\eta_1}} \omega^\mu \, , \nonumber \\
A_3^\mu & \equiv & \sqrt{\frac{\eta_3}{\eta_1}} \chi^\mu -
\sqrt{-\frac{\eta_2}{\eta_1}} \omega^\mu \, .
\end{eqnarray}
Under the action of the gauge group, $A_2$ and $A_3$ transform as
matter fields in the adjoint representation, while $A_1$ transforms as
a gauge field, due to the additional shift in $\hat{A}$. The inverse
transformations are given by
\begin{eqnarray}
\hat A^\mu & = & A_1^\mu - \sqrt{-\frac{\eta_2}{\eta_1}} A_2^\mu
+ \sqrt{\frac{\eta_3}{\eta_1}} A_3^\mu \, , \nonumber \\
\chi^\mu & = & \sqrt{-\frac{\eta_2}{\eta_1}} A_2^\mu -
\sqrt{\frac{\eta_3}{\eta_1}} A_3^\mu \, , \nonumber \\
\omega^\mu & = & \sqrt{\frac{\eta_3}{\eta_1}} A_2^\mu -
\sqrt{-\frac{\eta_2}{\eta_1}} A_3^\mu \, , \label{YMtrans}
\end{eqnarray}
as may be shown by using the sum rule Eq.~(\ref{eq:sr1}).
Substituting Eqs.~(\ref{YMtrans}) into Eq.~(\ref{YMaux}) is a
laborious but straightforward procedure.  Using
Eqs.~(\ref{eq:etadef1})--(\ref{eq:etadef3}) to express the parameters
$\eta_i$ in terms of masses $m_{2,3}$, and defining the unhatted field
strength $F_1^{\mu \nu}$ and covariant derivative $D^\mu$ as analogous
to Eqs.~(\ref{fs})--(\ref{cov}) with $\hat A^\mu
\! \to \! A_1^\mu$, one obtains the Lagrangian
\begin{equation}
{\cal L}_{\rm YM,\,LW} = {\cal L}_0 + {\cal L}_1 + {\cal L}_2 \, ,
\end{equation}
where the subscript indicates the power of $g$ that appears in the
coefficient of each gauge-invariant term.  The kinetic and mass terms
are contained in
\begin{eqnarray}
{\cal L}_0 & = & -\frac 1 2 \, {\rm Tr} \, F_1^{\mu \nu} F_{1 \mu \nu}
+ \frac 1 2 \, {\rm Tr} (D_\mu A_{2 \nu} - D_\nu A_{2 \mu} )^2 - \frac
1 2 \, {\rm Tr} ( D_\mu A_{3 \nu} - D_\nu A_{3 \mu} )^2
\nonumber \\ & & - m_2^2 \, {\rm Tr} A_2^\mu A_{2 \mu} + m_3^2
\, {\rm Tr} A_3^\mu A_{3 \mu} \, ,
\end{eqnarray}
from which one immediately sees that $A_1$ is massless $(m_1 \! = \!
0)$, and only $A_2$ has wrong-sign quadratic terms,
\begin{eqnarray}
\lefteqn{{\cal L}_1 = \frac {-ig}{m_3^2 - m_2^2} {\rm Tr} \left( F_{1
\mu \nu} \left[ m_3 A_2^\mu - m_2 A_3^\mu, \, m_3 A_2^\nu - m_2
A_3^\nu \right] \right)} \nonumber \\
& & + \frac{ig}{(m_3^2 - m_2^2)^{1/2}} \left\{ {\rm Tr} \, \left(
D_\mu A_{2 \nu} - D_\nu A_{2 \mu} \right) \left( 2m_3 \left[ A_2^\mu ,
A_2^\nu \right] - m_2 \left[ A_2^\mu, A_3^\nu \right] - m_2 \left[
A_3^\mu , A_2^\nu \right] \right) \right. \nonumber \\
& & \hspace{6.5em} \left. + {\rm Tr} \left( D_\mu A_{3 \nu} - D_\nu
A_{3 \mu} \right)
\left( 2m_2 \left[ A_3^\mu , A_3^\nu \right] - m_3 \left[ A_2^\mu,
A_3^\nu \right] - m_3 \left[ A_3^\mu , A_2^\nu \right] \right)
\right\} \, , \nonumber \\
\end{eqnarray}
and finally,
\begin{eqnarray}
\lefteqn{{\cal L}_2 = \frac{g^2}{2(m_3^2-m_2^2)^2}}  \nonumber \\
& & \times\Big\{ m_3^2 (4m_2^2 \! - 3m_3^2) {\rm Tr} \left[
A_2^\mu , A_2^\nu \right]^2 + 2 m_2^2 m_3^2 \,{\rm Tr} \left[ A_2^\mu
, A_2^\nu \right] \left[ A_{3\mu} , A_{3\nu} \right] + \, m_2^2
(4m_3^2 \! - 3m_2^2) {\rm Tr} \left[ A_3^\mu , A_3^\nu \right]^2
\nonumber \\
& & + 2m_2 m_3 (m_3^2 - 2m_2^2) {\rm Tr} \left[ A_2^\mu , A_2^\nu
\right] \left( \left[ A_{2 \mu} , A_{3 \nu} \right] + \left[ A_{3
\mu}, A_{2 \nu} \right] \right) \nonumber \\
& & + 2m_2 m_3 (m_2^2 - 2m_3^2) {\rm Tr} \left[ A_3^\mu , A_3^\nu
\right] \left( \left[ A_{2 \mu} , A_{3 \nu} \right] + \left[ A_{3
\mu}, A_{2 \nu} \right] \right) \nonumber \\
& & + (m_2^4 - m_2^2 m_3^2 + m_3^4) {\rm Tr} \left( \left[
A_2^\mu , A_3^\nu \right] + \left[ A_3^\mu , A_2^\nu \right] \right)
\left( \left[ A_{2 \mu} , A_{3 \nu} \right] + \left[ A_{3 \mu} ,
A_{2 \nu} \right] \right) \! \Big\} \, .
\label{eq:fourgauge}
\end{eqnarray}

While these expressions appear rather involved, they are substantially
simpler than they could be, owing to the sum rules
Eqs.~(\ref{eq:sr1})--(\ref{eq:sr3}).  Note that the decay $A_3 \! \to
\! A_2 A_1$ follows from the first term in ${\cal L}_1$ since
$m_3>m_2$.  In a complete theory, including fermions and Higgs fields,
decay channels open for $A_2$ as well.

\section{Fermions}\label{sec:fermions}

The next-to-leading-order HD Lagrangian for a chiral
fermion field $\hat{\phi}_L$ assumes the compact form
\begin{equation}
{\cal L}_{\rm HD, \, f} = \frac{1}{m_2^2 m_3^2} \overline{\hat \phi}_L
\left [ ( i \hat{\Dslash} )^2 - m_2^2 \right] \left [ ( i \hat{\Dslash}
)^2 - m_3^2 \right] i \hat{\Dslash} \hat \phi_L \ , \label{HDf}
\end{equation}
where $\hat\Dslash$ includes both the gauge bosons and their LW
partners.  This HD Lagrangian may be obtained from the equivalent
Lagrangian
\begin{eqnarray}
{\cal L}_{\rm f} & = & \overline{\hat \phi}_L i \hat{\Dslash} \hat
\phi_L - \overline{\chi}_R i \hat{\Dslash} \chi_R + \overline \psi_L
i \hat{\Dslash} \psi_L
+ ( \overline{\hat \phi}_L i \hat{\Dslash} \chi_L + {\rm h.c.} ) + (
\overline{\chi}_R i \hat{\Dslash} \psi_R + {\rm h.c.} ) \nonumber \\
& & + \frac{m_2 m_3}{m_2 + m_3} \left[ \overline{\chi}_R \left( \chi_L
+ \psi_L \right) + {\rm h.c.} \right] - (m_2 + m_3) \left(
\overline{\psi}_L \psi_R + {\rm h.c.} \right) \, . \label{faux}
\end{eqnarray}
The fields $\chi_L$ and $\psi_R$, which like $\hat{\phi}_L$ are Weyl
spinors transforming in the fundamental representation of the gauge
group, appear only linearly in Eq.~(\ref{faux}), and therefore may be
considered auxiliary.  Varying ${\cal L}_{\rm f}$ with respect to them
yields the constraints
\begin{eqnarray}
& & i \hat{\Dslash} \hat \phi_L + \frac{m_2 m_3}{m_2 + m_3}
\chi_R =0 \, , \label{fconst1} \\
&& i \hat{\Dslash} \chi_R - (m_2 + m_3) \psi_L =0 \, , \label{fconst2}
\end{eqnarray}
which may be substituted directly into ${\cal L}_{\rm f}$ to eliminate
all terms linear in $\chi_L$ and $\psi_R$, and also to re-express the
the remaining fields $\chi_R$, $\psi_L$ in terms of $\hat \phi_L$:
\begin{eqnarray}
\chi_R & = & -\frac{m_2 + m_3}{m_2 m_3} i \hat{\Dslash} \hat \phi_L
\, , \\ \psi_L & = & \frac{i \hat{\Dslash}}{m_2 + m_3}
\chi_R = -\frac{1}{m_2 m_3} (i \hat{\Dslash})^2 \hat \phi_L \, ,
\end{eqnarray}
where the final equality is obtained by substituting
Eq.~(\ref{fconst1}) into Eq.~(\ref{fconst2}).  It is straightforward
to check that these EOMs transform Eq.~(\ref{faux}) into the HD form
Eq.~(\ref{HDf}).

In order to obtain a Lagrangian in the LW form, we rewrite the three
left-handed fields $\hat \phi_L$, $\chi_L$ and $\psi_L$ in terms of
three new fields $\phi^{(1,2,3)}_L$, and the two right-handed fields
$\chi_R$, $\psi_R$ in terms of two new fields $\phi^{(2,3)}_R$:
\begin{eqnarray}
\phi^{(1)}_L & \equiv & \hat \phi_L + \chi_L \, , \nonumber \\
\phi^{(2)}_L & \equiv & \sqrt{-\frac{\eta_2}{\eta_1}} \, \chi_L -
\sqrt{\frac{\eta_3}{\eta_1}} \, \psi_L \, , \nonumber \\
\phi^{(3)}_L & \equiv & \sqrt{\frac{\eta_3}{\eta_1}} \, \chi_L -
\sqrt{-\frac{\eta_2}{\eta_1}} \, \psi_L \, ,
\end{eqnarray}
and
\begin{eqnarray}
\phi^{(2)}_R & \equiv & \sqrt{-\frac{\eta_2}{\eta_1}} \, \chi_R -
\left[ \sqrt{-\frac{\eta_2}{\eta_1}} + \sqrt{\frac{\eta_3}{\eta_1}}
\, \right] \psi_R \, , \nonumber \\
\phi^{(3)}_R & \equiv & \sqrt{\frac{\eta_3}{\eta_1}} \, \chi_R -
\left[ \sqrt{-\frac{\eta_2}{\eta_1}} + \sqrt{\frac{\eta_3}{\eta_1}}
\, \right] \psi_R \, .
\end{eqnarray}
The inverse transformations, whose simplification uses the sum
rule Eq.~(\ref{eq:sr1}), are
\begin{eqnarray}
\hat \phi_L & = & \phi^{(1)}_L - \sqrt{-\frac{\eta_2}{\eta_1}} \,
\phi^{(2)}_L + \sqrt{\frac{\eta_3}{\eta_1}} \, \phi^{(3)}_L \, ,
\nonumber \\
\chi_L & = & \sqrt{-\frac{\eta_2}{\eta_1}} \, \phi^{(2)}_L -
\sqrt{\frac{\eta_3}{\eta_1}} \, \phi^{(3)}_L \, ,
\nonumber \\
\psi_L & = & \sqrt{\frac{\eta_3}{\eta_1}} \, \phi^{(2)}_L -
\sqrt{\frac{-\eta_2}{\eta_1}} \, \phi^{(3)}_L \, ,
\end{eqnarray}
and 
\begin{eqnarray}
\chi_R & = & \left[ \sqrt{-\frac{\eta_2}{\eta_1}} +
\sqrt{\frac{\eta_3}{\eta_1}} \, \right] \left[ \phi^{(2)}_R -
\phi^{(3)}_R \right] \, , \nonumber \\
\psi_R & = & \sqrt{\frac{\eta_3}{\eta_1}} \, \phi^{(2)}_R -
\sqrt{\frac{-\eta_2}{\eta_1}} \, \phi^{(3)}_R \, .
\end{eqnarray}
Substituting these transformations into Eq.~(\ref{faux}) and using the
sum rules Eqs.~(\ref{eq:sr1})--(\ref{eq:sr3}) leads to a remarkable
set of simplifications.  Once the parameters $\eta_i$ are expressed in
terms of masses $m_2$, $m_3$, the LW fermion Lagrangian reads
\begin{equation}
{\cal L}_{\rm f , \, LW} = \overline{\phi}^{(1)}_L i \hat{\Dslash}
\phi^{(1)}_L - \overline{\phi}^{(2)} \! ( i \hat{\Dslash} - m_2 )
\phi^{(2)} + \overline{\phi}^{(3)} \! ( i \hat{\Dslash} - m_3 )
\phi^{(3)} \, ,
\end{equation}
where of course $\phi \equiv \phi_L \! + \phi_R$.  Note from the signs
of the terms that $\phi^{(2)}$ and $\phi^{(1),(3)}$ are negative- and
positive-norm states, respectively.  The HD, AF and LW Lagrangians for
a right-handed chiral fermion field $\hat{\phi}_R$ can be obtained
from those presented here by the exchange $R \leftrightarrow L$
throughout.  The results can then be applied immediately to any chiral
gauge theory (in particular, to the SM) without significant
modification.

\section{The Higgs Sector}\label{sec:higgs}

The discussion of the theory of a real scalar field in
Section~\ref{sec:scalarex} can be generalized in a straightforward way
to one of a complex scalar $\hat H$ that transforms in the fundamental
representation of a non-Abelian gauge group.  Let us first consider
the case in which the squared scalar mass is positive, $m^2_H >0$.
The HD Lagrangian may be written
\begin{equation}
{\cal L}_{\rm HD} = \hat D_\mu \hat{H}^\dagger \hat D^\mu \hat{H}
-m_H^2 \hat{H}^\dagger \hat{H}- \frac{1}{M_1^2} 
\hat{H}^\dagger (\hat D_\mu \hat D^\mu)^2 \hat{H} - \frac{1}{M_2^4}
\hat{H}^\dagger (\hat D_\mu \hat D^\mu)^3 \hat{H}  +
{\cal L}_{{\rm int}}(\hat{H})\,\,\,,
\end{equation}
where $m_H^2$, $M_1^2$ and $M_2^2$ are given by
Eqs.~(\ref{eq:lamsq})--(\ref{eq:msqds3}) with the identification
$m_\phi^2=m_H^2$.  The auxiliary field Lagrangian analogous to
Eq.~(\ref{eq:afstotal}) is
\begin{eqnarray}
{\cal L}_{{\rm AF}} &=& \frac{1}{\eta_1} \left\{ \hat D_\mu
\hat{H}^\dagger \hat D^\mu \hat{H} -m_1^2  \hat{H}^\dagger \hat{H}
- \left[  \chi^\dagger ( \hat D_\mu \hat D^\mu+m_1^2) \hat{H}
+ \mbox{ h.c.} \right] \right. \nonumber \\
&&+\left.(m_2^2-m_1^2)^{1/2}(m_3^2-m_1^2)^{1/2}(\chi^\dagger \psi +
\psi^\dagger \chi)
+ \hat D_\mu \psi^\dagger \hat D^\mu \psi - (m_2^2+m_3^2-m_1^2)
\psi^\dagger \psi \right\} \nonumber \\
&&+ {\cal L}_{{\rm int}}(\hat{H})\,\,\,,
\label{eq:afhiggs}
\end{eqnarray}
where $\psi$ and the auxiliary field $\chi$ also transform in the
fundamental representation. Again, one recovers the HD form of the
Lagrangian by applying the constraint equation obtained from varying
with respect to $\chi$.  The standard LW form of the theory is
obtained via field redefinitions identical to
Eqs.~(\ref{eq:redef1})--(\ref{eq:redef3}), with the relabelling
$\hat{\phi} \rightarrow \hat{H}$ and $\phi^{(i)} \rightarrow H^{(i)}$:
\begin{eqnarray}
{\cal L} &=&  - H^{(1)\dagger}(\hat D_\mu \hat D^\mu + m_1^2) H^{(1)}
+ H^{(2)\dagger} ( \hat D_\mu \hat D^\mu+m_2^2) H^{(2)} \\ \nonumber 
&&- H^{(3)\dagger} ( \hat D_\mu \hat D^\mu+m_3^2) H^{(3)}
+ {\cal L}_{{\rm int}} (\hat{H})  \,\,\,,
\end{eqnarray}
where 
\begin{equation}
{\cal L}_{{\rm int}}(\hat{H}) = {\cal L}\left( \sqrt{\eta_1} H^{(1)} -
\sqrt{-\eta_2} H^{(2)} + \sqrt{\eta_3} H^{(3)}\right) \,\,\,.
\end{equation}

In the SM, spontaneous symmetry breaking is ensured by $m_H^2 < 0$.
In this case it is more convenient to absorb the $m_H^2$ term into
${\cal L}_{{\rm int}}$:
\begin{equation}
{\cal L}_{{\rm HD}} =  {\cal L}_{{\rm HD}}(m_H^2 = 0) +
{\cal L'}_{{\rm int}}(\hat{H}) \,\,\, ,
\end{equation}
\begin{equation}
-{\cal L'}_{{\rm int}}(\hat{H}) \equiv \frac{\lambda}{4} \left(
\hat{H}^\dagger \hat{H}  - \frac{v^2}{2}\right)^2 \,\,\,,
\end{equation}
where $v$ is the Higgs vacuum expectation value.  The mass parameters
$m_2$ and $m_3$ are now determined by
\begin{equation}
M_1^2 = \frac{m_2^2 m_3^2}{m_2^2+m_3^2} \,\,\,\,\, \mbox{ and }
\,\,\,\,\, M_2^2=m_2 m_3  \,\,\,.
\end{equation}
The $m_H^2=0$ part of the Lagrangian is handled via the steps
described in Sec.~\ref{sec:scalarex}.  Using the $m_1^2=0$ values of
the $\eta_i$ parameters (and noting that $\eta_1=1$), one then finds
that the canonical LW form of the Higgs-sector Lagrangian is given by
\begin{eqnarray}
{\cal L} &=& \hat D_\mu H^{(1)\dagger} \hat D^\mu H^{(1)} - \hat D_\mu
H^{(2)\dagger} \hat D^\mu H^{(2)} + \hat D_\mu H^{(3)\dagger}
\hat D^\mu H^{(3)}+m_2^2 H^{(2)\dagger} H^{(2)} \\ \nonumber 
&&-m_3^2 H^{(3)\dagger} H^{(3)} + {\cal L'}_{{\rm int}}\left( H^{(1)}
- \sqrt{-\eta_2} H^{(2)} + \sqrt{\eta_3} H^{(3)}\right)  \,\,\,,
\label{eq:llp}
\end{eqnarray}
where the last term may be expanded
\begin{eqnarray}
-{\cal L'}_{{\rm int}} &=& \frac{\lambda}{4} \left( H^{(1)\dagger}
H^{(1)} - \frac{v^2}{2}\right)^2
+\frac{\lambda}{2} \left( H^{(1)\dagger} H^{(1)} - \frac{v^2}{2}
\right) \\ \nonumber
&& \times\left\{ \left[ H^{(1)\dagger}(\sqrt{-\eta_2} H^{(2)} +
\sqrt{\eta_3} H^{(3)}) 
+ \mbox{ h.c.} \right] + |\sqrt{-\eta_2} H^{(2)} + \sqrt{\eta_3}
H^{(3)}|^2 \right] \\ \nonumber
&&+ \frac{\lambda}{4} \left\{ \left[ H^{(1)\dagger}(\sqrt{-\eta_2}
H^{(2)} + \sqrt{\eta_3} H^{(3)}) 
+ \mbox{ h.c.} \right] + |\sqrt{-\eta_2} H^{(2)} + \sqrt{\eta_3}
H^{(3)}|^2 \right\}^2 \,\,\,.
\label{eq:lpalone}
\end{eqnarray}
In analogy to the minimal theory~\cite{GOW}, one may work in unitary
gauge, in which
\begin{equation}
H^{(1)} = \left(\begin{array}{c} 0 \\ \frac{1}{\sqrt{2}}(v+h_1)
\end{array} \right) , \,\,\,\,\,
H^{(2)} = \left(\begin{array}{c} h_2^+ \\ \frac{1}{\sqrt{2}}
(h_2 + i P_2) \end{array} \right) ,
\,\,\,\,\,
H^{(3)} = \left(\begin{array}{c} h_3^+ \\ \frac{1}{\sqrt{2}}
(h_3 + i P_3) \end{array} \right) ,
\end{equation}
where the fields $h_i$, $P_i$ and $h^+_i$ represent the scalar,
pseudoscalar and charged Higgs components, respectively.  Note that
the mass terms in Eq.~(\ref{eq:llp}) are given by
\begin{eqnarray}
{\cal L}_{{\rm mass}} &=& \frac{1}{2} m_2^2 \,
(2 h_2^- h_2^+ + h_2^2 + P_2^2)
-\frac{1}{2} m_3^2 \, (2 h_3^- h_3^+ + h_3^2 + P_3^2)  \\ \nonumber 
&&- \frac{1}{2} m^2 (h_1 - \sqrt{-\eta_2} h_2 +\sqrt{\eta_3} h_3)^2
\,\,\,,
\end{eqnarray}
with $m^2 = \lambda v^2/2$, indicating that the charged and
pseudoscalar Higgs masses are given directly by the parameters $m_2$
and $m_3$.  The neutral Higgs mass matrix, however, is off-diagonal;
the mass eigenstate basis is obtained via a transformation that
preserves the form of the neutral Higgs kinetic terms, which are
proportional to $diag(1,-1,1)$, in the basis ($h_1$, $h_2$, $h_3$).
Such transformation matrices can be found numerically, as was
demonstrated, for example, in Ref.~\cite{carleb}.  Using such a
numerical diagonalization, and the results presented here, one can
study the phenomenology of the Higgs sector like any other multi-Higgs
doublet extension of the SM\@. Derivation of the mass matrices of the
LW gauge bosons and fermions is straightforward using the field
redefinitions determined in this and the last two sections.

\section{Application: Divergence Cancellation} \label{divkill}

In this section we consider the cancellation of divergences in an $N=3$ SU($N_c$) 
gauge theory with a single complex scalar field in the fundamental representation.
This discussion generalizes the one appearing in Section~III of Ref.~\cite{GOW}, and provides
a number of explicit calculations using the LW form of the theory.   We also check that
one-loop quadratic divergences cancel when chiral fermions are present.

One can learn much about the divergences of the theory by considering the HD form of the Lagrangian 
in Landau gauge, where the $N=3$ gauge boson propagator scales as $p^{-6}$ at high energies ($p$ 
denotes a generic momentum).  The complex scalar propagator also scales as $p^{-6}$, while the 
Faddeev-Popov ghost propagator scales as $p^{-2}$. The salient issue is whether the derivatives 
at the new interaction vertices in the HD theory compensate for the additional momentum suppression 
in the propagators.  In the $N=3$ theory, a vertex with $n$ vectors scales as $p^{8-n}$, a vertex 
with two scalars and $n$ vectors as $p^{6-n}$, and one with two ghosts and one gauge field as $p$.  
The steps for constructing the superficial degree of divergence, $d$ are identical to those 
discussed in Section~III of Ref.~\cite{GOW}, so we do not repeat them.  The result in the $N=2$ theory,
\begin{equation}
d=6-2\,L-E-E'-2\,E_g  \,\,\,\,\,\,\,\,\,\, (N=2) \, ,
\end{equation}
becomes
\begin{equation}
d=8-4\,L-E-E'-3\,E_g \,\,\,\,\,\,\,\,\,\, (N=3) \, ,
\end{equation}
where $L$ is the number of loops, $E$ is the number of external scalar lines, $E'$
is the number of external vector lines, and $E_g$ is the number of external ghosts.
[For arbitrary $N$, one finds $d=2(N+1)-2(N-1)L-E-E'-N\,E_g$.]  For the gauge boson and
complex scalar self-energies, $d=6-4\,L$;  the divergences are at most quadratic and occur at
no higher than one loop.  

In the case of the gauge boson self-energies, the cancellation of the potential quadratic divergence 
is a consequence of gauge invariance, as in the $N=2$ theory~\cite{GOW}. Amplitudes in the HD theory 
satisfy a Ward identity, which implies that the 1-particle irreducible two-point function for  
$\hat{A}$ must be of the form $(q^2 g_{\mu\nu}-q_\mu q_\nu)$ times a dimensionless function of the 
regulator scale and the external momentum $q^2$.  A straightforward power counting of HD Lagrangian
mass parameters shows that they only multiply the divergent parts of the possible one-loop diagrams in 
dimensionless ratios, so that the divergence is at most logarithmic.  An equivalent calculation in the 
LW form of the Lagrangian is possible but prohibitive in theories with $N>2$ due to the proliferation of 
gauge boson self-interactions [see, for example, Eq.~(\ref{eq:fourgauge})].  If a chiral fermion is 
added to the $N=3$ theory, one finds that the fermion-vector coupling scales as $p^4$, the 
fermion/two-vector coupling scales as $p^3$, and the fermion propagator as $p^{-5}$.  It follows 
immediately that the one-loop fermion contributions to the gauge boson self-energy have $d=2$; the 
quadratic divergence cancels for the same reason as in the purely bosonic loop diagrams.

In the case of the complex scalar, on the other hand, it is straightforward to show 
the cancellation of one-loop divergences in the LW form of the theory.  We present the explicit 
calculation below as an illustration of the formalism.
\begin{figure}
\includegraphics[scale=.5]{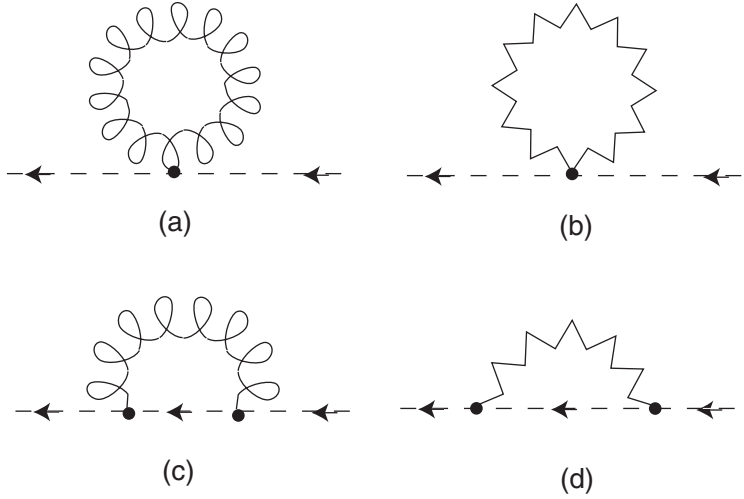}
\caption{\label{fig:newfig} Diagrams that contribute to the mass renormalization of 
the complex scalars.  The dashed lines refer to the field $H^{(i)}$, for $i=1,2$ or $3$. The 
curly lines represent the ordinary gauge field $A^{(1)}$; the zigzag lines represent its LW 
partners $A^{(2)}$ or $A^{(3)}$.}
\end{figure}

\subsection{The ordinary scalar}
We first consider the mass renormalization of the ordinary complex scalar field $H_1$. The $\eta_i$ 
shown in the formulae below are functions Eqs.~(\ref{eq:etadef1})--(\ref{eq:etadef3}) of the gauge 
boson masses $m_1=0$, $m_2$ and $m_3$.  We make the same assumptions as Ref.~\cite{GOW}, that the 
scalar potential is vanishing so that the ordinary scalar is massless, and work in Feynman gauge. 
Equations~(32a)-(32d) in Ref.~\cite{GOW} generalize as follows:
\begin{eqnarray}
-i \Sigma_a(0) &=& g^2 C_2(N_c) \int \frac{d^nk}{(2\pi)^n} \frac{n}{k^2} \, ,
\label{eq:ordA} \\
\-i \Sigma_b(0) &=& -g^2 C_2(N_c) \int \frac{d^nk}{(2\pi)^n} \left[
\left( \frac{n-1}{k^2-m_2^2}-\frac{1}{m_2^2}\right)\left(-\frac{\eta_2}{\eta_1}\right) \right.
\nonumber \\
&&-\left.\left( \frac{n-1}{k^2-m_3^2}-\frac{1}{m_3^2}\right)\left(\frac{\eta_3}{\eta_1}\right)\right] \, ,
\label{eq:ordB} \\
-i \Sigma_c(0) &=&  -g^2 C_2(N_c) \int \frac{d^nk}{(2\pi)^n} \frac{1}{k^2} \, ,
\label{eq:ordC} \\
-i \Sigma_d(0) &=& -g^2 C_2(N_c) \int \frac{d^nk}{(2\pi)^n} \left[ \frac{1}{m_2^2}\left(
-\frac{\eta_2}{\eta_1}\right) - \frac{1}{m_3^2}\left(\frac{\eta_3}{\eta_1}\right)\right] \, .
\label{eq:ordD}
\end{eqnarray}
These results correspond to the diagrams shown in Fig.~\ref{fig:newfig}. The cancellation 
of quartic divergences [between Eqs.~(\ref{eq:ordB}) and (\ref{eq:ordD})] is obvious by 
inspection.  The quadratic divergence originates from
\begin{equation}
\frac{n}{k^2}+\frac{n-1}{k^2}\left(\frac{\eta_2+\eta_3}{\eta_1}\right)-\frac{1}{k^2} \,\,\,,
\label{eq:qdivcancel}
\end{equation}
where the terms are the $k^2 \gg m_i^2$ limits of the integrands of Eqs.~(\ref{eq:ordA}), (\ref{eq:ordB}) 
and (\ref{eq:ordC}), respectively.  This quantity vanishes because $\eta_1+\eta_2+\eta_3=0$.  
Hence, the ordinary scalar mass remains logarithmically divergent, as in the $N=2$ theory.

\subsection{The Negative-Norm LW scalar}
The normal scalar discussed in the last subsection has two LW partners in the $N=3$ theory.  We
first consider the shift in the pole mass of the lighter, negative-norm state, whose mass we
denote by $m_{H_2}$. Equations (33a)-(33d) in Ref.~\cite{GOW} generalize as follows:
\begin{eqnarray}
-i \Sigma_a(m^2_{H_2}) &=& -g^2 C_2(N_c) \int \frac{d^nk}{(2\pi)^n} \frac{n}{k^2} \, ,
\label{eq:nnormA} \\
-i \Sigma_b(m^2_{H_2}) &=& g^2 C_2(N_c) \int \frac{d^nk}{(2\pi)^n} \left[
\left( \frac{n-1}{k^2-m_2^2}-\frac{1}{m_2^2}\right)\left(-\frac{\eta_2}{\eta_1}\right) 
\right. \nonumber\\
&&-\left.\left( \frac{n-1}{k^2-m_3^2}-\frac{1}{m_3^2}\right)\left(\frac{\eta_3}{\eta_1}\right)\right] \, ,
\label{eq:nnormB} \\
-i \Sigma_c(m^2_{H_2}) &=& g^2 C_2(N_c) \int \frac{d^nk}{(2\pi)^n} \left[\frac{1}{k^2-2p\cdot k}
+\frac{4 m^2_{H_2}-4 p\cdot k}{k^2 (k^2-2 p \cdot k)}\right] \, ,
\label{eq:nnormC} \\
-i \Sigma_d(m^2_{H_2}) &=& g^2 C_2(N_c) \int \frac{d^nk}{(2\pi)^n} \left[ \left(
\frac{1}{m_2^2} - \frac{4 m^2_{H_2}-2 p\cdot k}{(k^2-m_2^2)(k^2-2 p \cdot k)}\right)
\left(-\frac{\eta_2}{\eta_1}\right) \right.\nonumber \\
& &-\left.\left( \frac{1}{m_3^2} - \frac{4 m^2_{H_2}-2 p\cdot k}{(k^2-m_3^2)(k^2-2 p \cdot k)}\right)
\left(\frac{\eta_3}{\eta_1}\right)\right] \, .
\label{eq:nnormD}
\end{eqnarray}
Terms manifestly odd in $k$ have been dropped. Quartically divergent terms clearly cancel between 
Eqs.~(\ref{eq:nnormB}) and (\ref{eq:nnormD}).  Quadratic divergences are found in 
Eqs.~(\ref{eq:nnormA}), (\ref{eq:nnormB}) and (\ref{eq:nnormC}), but again in a combination
proportional to $\eta_1+\eta_2+\eta_3=0$. Thus, quadratic divergences cancel between 
diagrams and only a logarithmic divergence remains.

\subsection{The Positive-Norm LW scalar}
The on-shell self-energies of the heavier, positive-norm LW scalar (with mass $m_{H_3}$)
may be obtained from Eqs.~(\ref{eq:nnormA})--(\ref{eq:nnormD}) by replacing 
$m_{H_2} \rightarrow m_{H_3}$, and by flipping the overall sign of these results.
The sign flip originates from the change in sign of the $H_3$ quadratic terms relative to those
of $H_2$.  In the $a$ and $b$ diagrams, the sign flip originates from the opposite sign of 
the two-scalar/two-gauge vertex; in the $c$ and $d$ diagrams, it originates from sign changes at 
each vertex and in the scalar propagator.  These modification do not alter the cancellation of 
divergences between diagrams, so that the positive-norm LW scalar mass also receives only logarithmic 
corrections.

\subsection{Yukawa couplings}

If chiral fermions are present in the theory, then one may also consider the effect of
Yukawa couplings like
\begin{equation}
{\cal L} = \lambda\,\left( \bar{\hat{\phi}}_L \hat{H} \hat{\psi}_R + \mbox{ h.c.}\right)\, ,
\end{equation}
where $\hat{\phi}_L$ transforms in the fundamental representation, while $\psi_R$ is
a singlet.  Letting $\eta_i$ refer to the LW mass spectrum of $\phi_L^{(i)}$ and
$\eta_i'$ to that of $\psi_R^{(i)}$, it is easy to see that the quadratically
divergent part of the one-fermion loop contribution to the complex scalar self-energy
is proportional to 
\begin{equation}
\left(1+\frac{\eta_2}{\eta_1}+\frac{\eta_3}{\eta_1}\right)
\left(1+\frac{\eta_2'}{\eta_1'}+\frac{\eta_3'}{\eta_1'}\right)
\end{equation}
which vanishes since $\eta_1+\eta_2+\eta_3=0$ (and similarly for the $\eta_i'$),
again confirming that the quadratic divergences are cancelled at one loop.

\section{Conclusions}\label{sec:concl}
The Lee-Wick Standard Model provides a new theory that is interesting
from both the formal field-theoretical and phenomenological points of
view.  Its means of solving the hierarchy problem, by cancelling the
leading divergences of loop diagrams between each particle and a
partner of the same statistics and quantum numbers but carrying
wrong-sign kinetic and mass terms, is innovative and worthy of
detailed study.
 
To this end, we have developed the generalization of the theory to
allow each particle two LW partners.  Since the original Lee-Wick
Standard Model~\cite{GOW} involves higher-derivative quadratic terms
of $O(p^4)$ in momentum space (for the bosonic fields), our theory
necessarily includes terms of $O(p^6)$.  Referring to the number of
poles in the two-point function, we name these the $N=2$ and $N=3$
Lee-Wick theories, respectively.  We note that there is no impediment,
in principle, that prevents the generalization of our approach to
theories with $N>3$.
 
The recasting of HD theories in terms of fields satisfying low-order
equations of motion (the Ostrogradsky method for reducing high-order
differential equations to a recursive system of low-order ones, as
applied to quantum field theory) was developed decades ago by Pais and
Uhlenbeck~\cite{pu}.  The results presented here are new in a number
of significant respects.  First, we supply the prescription for
rewriting a viable $N=3$ HD theory in terms of an equivalent AF theory
containing no terms of dimension higher than four; the $N \! = \! 2$
case was developed of course by Grinstein {\it et al.} in
Ref.~\cite{GOW}.  Such auxiliary fields provide constraints that are
exact at the quantum level, and once imposed, exactly reproduce the HD
Lagrangian.  On the other hand, the auxiliary fields may be rewritten
in terms of a set of fields whose quadratic terms are canonical, up to
overall signs, and whose couplings are intricately intertwined.  For
$N \! = 3$, these fields consist of the original particle, one
negative-norm and one positive-norm LW partner; the three fields
together conspire to cancel the quadratic divergences in the
theory. Notably, our $N\!=\!3$ analysis includes non-Abelian chiral
gauge theories, with or without spontaneous symmetry breaking, topics
that were not addressed in the ancient literature on nonlocal
Lagrangians.
 
We have successfully developed this construction, with minor
variations, in theories with real scalars, fermions, gauge bosons, and
complex scalars, and allowing for spontaneous symmetry
breaking.  One concludes that the entire Standard Model may be easily
embedded in an $N \! = 3$ LW theory, a possibility that offers an
abundant new wellspring for future studies of the formal properties
and phenomenology of the model.
\vspace{-0.5em}
\section*{Acknowledgments}
\vspace{-0.5em}
This work was supported by the NSF under Grant Nos.\ PHY-0456525 and
PHY-0757481 (CDC) and PHY-0757394 (RFL).

\end{document}